\documentclass[aps,pra,preprint,superscriptaddress]{revtex4}

\usepackage{graphicx}
\usepackage{dcolumn}
\usepackage{bm}
\usepackage{amsmath}
\usepackage{amssymb}
\usepackage{latexsym}
\usepackage{epsfig}
\usepackage{amsbsy}
\usepackage{array}
\usepackage{amssymb}
\usepackage{setspace}
\usepackage{bm}

\def\sint{\ifmmode{- \!\!\!\!\!\! \int}
    \else{\hbox{$- \!\!\!\! \int \ $}}\fi}

\begin{document}


\title{ Probing the dynamic interference in molecular high-order harmonic generation through Bohmian trajectories
}

\author{Yang Song}
\affiliation{Institute of Atomic and Molecular Physics, Jilin
University, Changchun 130012, China}
\author{Bing-Bing Wang}
\affiliation{Beijing National Laboratory for Condensed Matter
Physics, Institute of Physics, Chinese Academy of Sciences,
Beijing 100190, China}
\author{Fu-Ming Guo}
\author{Su-Yu Li}
\author{Xue-Shen Liu}
\affiliation{Institute of Atomic and Molecular Physics, Jilin
University, Changchun 130012, China}

\author{ Ji-Gen Chen}
\email[]{tz501@yahoo.cn}
 \affiliation{Department of Physics and
Materials Engineering, Taizhou University, Taizhou 318000, China}
\author{Si-Liang Zeng }
\email[]{zeng siliang@iapcm.ac.cn} \affiliation{Science and
Technology Computation Physics Laboratory, Institute of Applied
Physics and Computational Mathematics, Beijing 100088, China}
\affiliation{Data Center for High Energy Density Physics,
Institute of Applied Physics and Computational Mathematics,
Beijing 100088, China}
\author{Yu-Jun Yang}
\email[]{yangyj@jlu.edu.cn} \affiliation{Institute of Atomic and
Molecular Physics, Jilin University, Changchun 130012, China}


\begin{abstract}
By using Bohmian trajectory method, we investigate the dynamic
interference in diatomic molecular high-order harmonic generation
progress. It is demonstrated that the main characteristics of the
molecular harmonic spectrum can be well reproduced by only two
Bohmain trajectories which are located respectively at the two ions.
This is because these two localized trajectories can receive and
store the whole collision information coming from all of the other
recollision trajectories. Therefore, the amplitudes and frequencies
of these two trajectories represent the intensity and frequency
distribution of the harmonic generation. Moreover, the interference
between these two trajectories shows a dip in the harmonic spectrum,
which indicates the molecular structure information.

\end{abstract}

\pacs{ 32. 80. Rm, 42. 50. Hz}

\maketitle

When atoms and molecules are irradiated by an intense laser pulse,
 high-order harmonics of the laser's frequency may be generated~\cite{1,2,3,4}.
  The mechanism of high-order harmonic generation (HHG) is the recombination
  of the ionized electron with its parent ion in the laser field~\cite{5,6}.
  HHG offers us a new way to detect the microscopic process with sub-angstrom spatial
and attosecond temporal resolution~\cite{7,8,9}. HHG can also be
applied to imaging the molecular orbitals~\cite{10,11,12}, probing
the electronic or nuclear dynamical behavior with attosecond
resolution~\cite{13,14,15}.

The response of single atom or molecule of HHG has been studied
theoretically by using numerical solution of time-dependent
schr\"{o}dinger equation (TDSE)~\cite{16,17}. The TDSE calculation
can provide an accurate result to simulate the experiment
measurement. However, it is difficult to extract the dynamic
information of the HHG process from the time-dependent wavefunction
and to make clear the physical mechanism behind the process. To
overcome this difficulty, one can adopt the Bohmian trajectory(BT)
scheme. Recently, this scheme has been used in the research on the
interaction between matter and strong field. For example, by using
this method, Takemoto {\em et al.}~\cite{18} investigated the
attosecond electron dynamics of molecular ion, and Lai {\em et
al.}~\cite{19} studied the correspondence between the quantum and
classical processes in the strong field ionization. Furthermore,
this scheme has been successfully applied to investigate the atomic
HHG processes~\cite{19,20,21}. It is found that the calculation of
the atomic HHG by solving the TDSE agrees well with the result by
the BT scheme~\cite{21}.

In this work, the BT method is applied to study the dynamic interference process
in the molecular HHG processes. For the HHG of small linear molecules, there
is the minima in the harmonic spectra due to the interference of many atomic
centers~\cite{22,23,24,25,26,27}. This minima in the molecular HHG spectrum was
 firstly predicted theoretically by Lein {\em et al.}~\cite{16,17} and confirmed
 experimentally  by Kanai {\em et al.}~\cite{23}. The interference structure changes
 with the internuclear distance, the symmetry of molecular orbit, and
multi-orbit effect~\cite{27,28}. Thus it is necessary to make clear
the physical mechanism behind the interference structure in the
molecular HHG spectrum.

To understand the interference mechanism, we investigate the HHG of
H$_2^+$ molecular ion by using the BT scheme. It is found that the
HHG can be reproduced qualitatively by using only two BTs whose
initial positions are located at the two atomic centers respectively
in the molecular ion. More importantly, the accurate interference
structure in the molecular HHG spectrum can be illustrated through
the dynamic analysis of the Bohmian particles's acceleration. These
Bohmian particles play the role as an 'antenna' which can well
reflect the dynamic information about the electron in the molecule.

\section{Theoretical methods}  

We take the electron probability density as a fluid, and its flow can be analyzed
 by the Bohmian mechanics of many Bohmian particles whose
trajectories are guided by the quantum wavefunction~\cite{17,29,30,31}.
 The quantum wavefunction $\psi (x,t)$ was obtained from the numerical solution
of the one-dimensional TDSE (atomic units are used throughout this
paper, unless otherwise stated): $ i{\partial \psi (x,t)}{/{\partial
t}}=H \psi (x,t)$, where
$H=-{\partial}^{2}/(2\partial{x}^{2})+V(x,R,t)$ is the Hamtonian of
the H$_2^+$ molecular ion. Here $R$ is the distance between two
protons, and $x$ is electronic coordinate in the molecular frame.
Under dipole approximation and in length gauge, the time-dependent
potential is $V(r,R,t)=V(x,R)+E(t)\cdot x$, where the soft-core
potential $V(x,R)= - 1/\sqrt {{{(x - R/2)}^2} + \alpha} - 1/\sqrt
{{{(x + R/2)}^2} +  \alpha}$  is chosen to represent the interaction
between the electron and the nuclei, where $\alpha$ is the soft core
parameter. The laser-electron interaction potential is $E(t)\cdot
x$, where the laser's electric field is $E{\rm{(}}t{\rm{) =
}}{E_0}f{\rm{(}}t{\rm{)}}\sin {\rm{(}}\omega t{\rm{)}}{\rm{}}$ with
$\omega$ and $E_0$ being the frequency and peak amplitude of the
laser pulse, respectively. The envelope of the laser's electric
field is $f(t) = \exp [- 4\ln 2{\left( {(t -T/2)/\sigma }
\right)^2}]$ with the total width being $T=400$.

The TDSE are solved numerically using symmetrically splitting fast Fourier transformation scheme~\cite{32,33}. By using the time-dependent wavefunction, the BTs $\left\{x_{k}(t)|k=1,...,N_{Tra}\right\}$ are propagated by solving the equation:

\begin{align}{\dot{x}_{k}}(t) = {\mathop{\rm Im}\nolimits} \left[ {\frac{1}{{\psi (x,t)}}\frac{\partial }{{\partial x}}\psi (x,t){|_{x = {x_k(t)}}}} \right].\end{align}

The initial positions of the Bohmian particles are selected by the
electron probability density of the ground state of H$_2^+$, and for each
trajectory we assign the same weight. The following positions of
these Bohmian particles are calculated from the integration of Eq.
(1)~\cite{28}. Furthermore, one can obtain the acceleration
$a_{k}(t)=\ddot{x}_{k}(t)$ of the Bohmian particles by taking the derivative of Eq.~(1). It should be noticed, in classical viewpoint, that the square of the particle's acceleration is proportional to the instantaneous power of the
radiation. Hence the harmonic spectrum of each particle can be obtained by using the Fourier transformation scheme:
\begin{align}P_{k}(\omega ) = {\left| {\frac{1}{{{t_f} - {t_i}}}\frac { 1 }{\omega^{2}  } \int_{{t_i}}^{{t_f}} {a_{k}(t){e^{ - i\omega t}}dt} } \right|^2}{\rm{,}}\end{align}
where $t_i$ and $t_f$ are the initial and final instants of the
laser pulse, respectively. The whole HHG of the molecular ion can be
calculated from the Fourier transform of the acceleration of the
total Bohmian trajectories :

\begin{align}P(\omega ) = {\left| {\frac{1}{{{t_f} - {t_i}}}\frac { 1 }{\omega^{2}  } \int_{{t_i}}^{{t_f}} {a_{\rm BT}(t){e^{ - i\omega t}}dt} } \right|^2}{\rm{,}}\end{align}
where $a_{\rm BT}(t)=\sum _{k=1}^{N_{\rm Tra}}{a_{k}(t)/N_{\rm Tra}
}$.

For the purpose of comparison, we also calculated the HHG from TDSE
using the time dependent acceleration dipole:
\begin{align}P(\omega ) = {\left| {\frac{1}{{{t_f} - {t_i}}}\frac { 1 }{\omega^{2}  } \int_{{t_i}}^{{t_f}} {a(t){e^{ - i\omega t}}dt} } \right|^2}{\rm{,}}\end{align}
where

\begin{align}a(t) =  \left\langle {\psi (x,t)\left| {\frac{{dV(x,R)}}{{dx}} - E(t)} \right|\psi (x,t)} \right\rangle .\end{align}

\section{Results and discussions}  
In order to explore the generation mechanism of the minima structure in the
H$_2^+$ molecular HHG spectrum, we first calculate the BTs using Eq. (1).
 In the insert of Fig. 1 (a), we present the initial position of the the Bohmian particles
 that are sampled from the electronic density distribution of the ground state,
 which is calculated from the H$_2^+$ molecular potential with $ \alpha=0.4$ and $R=8.5$.
 In our calculation, the laser's frequency is $\omega=0.057$, and the peak amplitude of the
 laser's electric field is $E_0=0.1$, where the shape of the laser's electric field is shown
 in the insert of Fig.~1(b). The Bohmian trajectories are shown in Fig,~1(a), where we may find
  that most of these trajectories are still located at around the two atomic nuclei and few trajectories are ionized after the laser
   pulse. In order to clearly
   distinguish the trajectories, we only present 400 trajectories in
   the Fig.1 (a).
    One may ask: can these un-ionized trajectories play a role in HHG? If the answer is yes,
    then how these bound BTs play roles in HHG? To answer these questions, we select two typical
    BTs which are initially located at -$R/2$ (BT(N)) and  $R/2$ (BT(P)) respectively. The trajectory BT(N)
    is expressed in cyan curve, and BT(P) is expressed in green line in Fig.~1(a).
     We also present a recollision trajectory (the red curve) in Fig.~1(a). For the purpose of comparison, in Fig. 1(b), we present
the time-evolution picture of electronic probability density
distribution by solving TDSE. We can find that the time-dependent
electronic probability density agrees well with the evolution of the
BTs. Under the dominance of the laser's electric field, the
ionization mainly occurs at two moments ($t=200$ and $t=250$) which
correspond to the peak positions of two half cycles of the laser
electric field. Moreover, only for the ionization at about $t=200$,
the ionized electron has chance to come back to the nucleus (red
line in Fig. 1(a)).

Using the two selected trajectories BT(N) and BT(P), we calculate the
corresponding harmonic spectra by Eq.~(2). The results are
presented in Fig. 2(a), where the solid black line is the harmonic
 spectrum calculated from the trajectory BT(N), and the dotted red line is
 the harmonic spectrum calculated from the trajectory BT(P). It can be seen from Fig.~2(a)
 that the two harmonic spectra have almost same intensities but different cutoffs. For the low order region of the harmonic spectra,
 their harmonic structures are similar with each other. As the harmonic order is larger
 than 30, the difference between these two harmonic
 spectra increases with the harmonic order. As we calculate the total harmonic
  spectrum by summing up the contributions of these two trajectories, a clear minimum
  at about 33rd harmonic order appears in the HHG spectrum caused by the interference between these
  two trajectories, as shown by a blue arrow in Fig.~2(b). This result qualitatively agrees with the
  results of the numerical calculation of TDSE and that by summing up the contributions of all the
   Bohmian trajectories ($N_{Tra}=50000$), as shown by the dash-dotted green line and dotted red line in Fig. 2(b), respectively.

In order to explain why only two BTs can simulate the structure of the whole HHG spectrum,
we investigate the dynamic behaviors of the two corresponding
Bohmian particles and their accelerations. The time evolution of BT(N) and BT(P)
 is presented in Fig.~3(a). From Fig.~3(a) we can see that two particles oscillate
 with time, whose oscillation amplitudes are proportional to the change of electric
field amplitude. The behaviors of the two Bohmian particles are alike at
the beginning of the laser pulse and exhibit the difference at the moment about $t=200$.
 At this moment, the trajectory BT(P) appears a fast oscillation whose frequency is about
 10 $\omega$, while BT(N) moves smoothly. Furthermore, at about moment $t=250$, BT(P) moves
 smoothly while BT(N) oscillates. These oscillation mainly contributes to the low order harmonics in the HHG spectrum, and its
generation mechanism has been explained by Wang {\em et al}~\cite{34}.

Because the radiation of a particle is proportional to the module
square of its acceleration~\cite{35},
we present the corresponding
acceleration of two particles in Fig.~3(b). It can be clearly seen
from Fig.~3(b) that around the instant 200, only BT(P) oscillates
with a non-negligible amplitude whose frequency is about 10
$\omega$. However, in the period of $t=230-300$, a fast oscillation
with high amplitude appears for both Bohmian particles with almost
the same amplitude. In order to clearly investigate the dynamic
interference profile of the HHG, we focus on the accelerations of
the two Bohmian particles in the period of $t=230-300$, as is shown
in Fig. 3(c). From this figure one can see that the dipole
accelerations of two particles have similar magnitude and
oscillation frequency at every instant. However, their relative
phase changes with time. In the regions 'A', 'C' and 'E', their
accelerations exhibit opposite phases, while in the regions 'B' and
'D' , their accelerations exhibit same phases. From the viewpoint of
the quantum interference, we may predict that when the two
oscillations have the same phases, the corresponding harmonic
emission should be coherently enhanced; On the contrary, when the
two oscillations have the opposite phases, the corresponding
harmonic emission should be coherently reduced. To confirm our
prediction, we perform the time-frequency analysis using wavelet
transform for the Bohmian particles:
\begin{align}{A_\omega }({t_0},\omega ) = \int_{{t_i}}^{{t_f}} {{a_k}(t){w_{{t_0},\omega }}(t)dt} {\rm{,}}\end{align}

where the wavelet kernel is ${w_{{t_0}{\rm{,}}\omega }}{\rm{ =
}}\sqrt \omega  W{\rm{(}}\omega {\rm{(}}t{\rm{ - }}{t_0}{\rm{))}}$,
and Morlet wavelet\cite{36,37} is used in this work.

It should be noticed that the
 dynamic emission profile of the harmonic can be clearly observed from
  the corresponding acceleration of the Bohmian particle. In Fig. 4(c),
   we present the time-frequency profile of the harmonic generated from
   the summation of BT(P) and BT(N). For the higher order harmonics,
    in the period of $t=230-300$, the interference pattern between these trajectories can be observed. Compared with time-dependent
acceleration in Fig. 3(c), we find that, in time region of 'A', 'C'
and 'E', where the two particles are out of phase, the harmonic
intensity is low. On the contrary, in the period of 'B' and 'D', the
Bohmian particles' accelerations are in phase, and hence the
corresponding harmonic emission has higher intensity. Thus one can
clearly observe the dynamic interference profile of molecular HHG
through the analysis of BTs. The time frequency profiles of the
harmonic generated from two Bohmian particles are shown in Fig.~4(a)
and (b).  In Fig. 4(d), we calculate the time-frequency profile from
the dipole calculated from the TDSE. Comparing Fig.~4(d) with (c),
we can find that, in the high energy part, this dynamic profile of
HHG agrees well with the result by the coherent summation of the two
Bohmian trajectories.

The reason why the main characteristics of the molecular harmonic
spectrum can be generated by only two Bohmain trajectories which are
 located respectively at the two ions is that the two localized trajectories
 can receive and store the whole collision information coming from all of the
  other recolliding trajectories. The motion of Bohmian particle is not affected by
 classical force but also affected by the quantum force \cite{21}. Therefore, using the information about the two BTs, we can detect the
recolliding process.

\section{Conclusion}  

In conclusion, utilizing Bohmian trajectory scheme, the dynamic
coherent process of the molecular high-order harmonic generation is
investigated in this paper. Through analyzing the temporal
characteristics of Bohmian trajectories, we found that the minima in
the molecular HHG spectrum is attributed to the interference of the Bohmian trajectories located at the two centers of molecule. The Bohmian trajectory scheme can clearly reflect the dynamic process of harmonic interference, and can be
taken as a detector to explore the dynamic process of complicated
molecular system.

\begin{acknowledgments}
This work was supported by the National Basic Research Program of
China (973 Program) 2013CB922200, the National Natural Science
Foundation of China under Grants No. 11274141, No.11034003, No.
61275128, No. 10904006, No.11247024 and No. 11274001, and science
and Technology Funds of China Academy of Engineering Physics under
Grant No.2011B0102026. Y.-J. Yang acknowledge the High Performance
Computing Center of Jilin University for supercomputer time.
\end{acknowledgments}


\begin{thebibliography}{apsrev4-1}

\bibitem{1} A. McPherson, G. Gibson, H. Jara, U. Johann, T. S. Luk,
I. A. McIntyre, K. Boyer and C. K. Rhodes, J. Opt. Soc. Am. B
\textbf{4,} 595 (1987).

\bibitem{2} M. Ferray, A. L'Huillier, F. Li, L. Lompre, G. Mainfray
and C. Manus, J. Phys. B \textbf{21,} L31(1998).

\bibitem{3} T. Popmintchev, M. C. Chen, D. Popmintchev, P. Arpin, S. Brown, S. Alisaukas,
G. Andriukaitis, T. Balciunas, O. D. Mucke, A. Pugzlys, A.
Baltusaka, B. Shim, S. E. Schrauch,
 A. Gaeta, L. Plajs, A. Becker, A. Jaron-Backer, M. M. Murnane and H. C. Kapteyn, Science \textbf{336,}, 1287 (2012).
\bibitem{4} M. Lein, J. Phys. B \textbf{40,} R135(2007).

\bibitem{5}  P. B. Corkum, Phys. Rev. Lett., \textbf{71,} 1994 (1993).

\bibitem{6} M. Lewenstein, P. Balcou, M. Y. Ivanov, A. L'Huillier and P. B. Corkum, Phys. Rev. A, \textbf{49,} 2117(1994).
\bibitem{7} P. Salieres, A. Maquet, S. Haessler, J. Caillat and R.
Taibeb, Rep. Prog. Phys., \textbf{75}, 062401(2012).

\bibitem{8} E. Goulielmakis, M. Schultze, M. Hofstetter, V. S.
Yakovlev, J. Gagnon, M. Uiberacker, A. L. Aquila, E. M. Gulikson, D.
T. Attwood, R. Kienberger, F. Krausz, and U. Kleineberg, Science
\textbf{302,} 1614(2008).
\bibitem{9}F. Krausz, M. Ivanov, Rev. Mod. Phys. \textbf{81,} 163 (2009).

\bibitem{10} J. Itatani, J. Levesue, D. Zeidler, H. Niikura, H. Pepin, J. C. Kieffer, P. B. Corkum, and D. M. Villeneuve, Nature (London), \textbf{432,} 867(2004).
\bibitem{11} B. McFarland, J. Farrell, P. Bucksbaum and M. Guhr,
Science \textbf{322,} 1232 (2008).
\bibitem{12} S. Haessler, J. Caillat and P. Salieres, J. phys. B
\textbf{44,} 203001(2011).

\bibitem{13} W. Li, X. Zhou, R. Lock, S. Patchkovskii, A. Stolow, H.
C. Kapteyn, and M. M. Murnane, Science \textbf{322,} 1207 (2008).
\bibitem{14} O. Smimova, Y. Mairesse, S. Patchkovskii N. Dudovich,
D. Villeneuve, P. Corkum, and M. Yu. Ivanov, Nature \textbf{460,}
972 (2009).
\bibitem{15} H. J. Worner, J. B. Bertrand, D. V. Kartashov, P. B.
Corkum, and D. M. Villeneuve, Nature \textbf{446,} 604(2010).

\bibitem{16}  M. Lein, N. Nay, M. R. Velotta, J. P. Marangos and P. L. Knight, Phys. Rev. Lett.
\textbf{88,} 183903 (2002).

\bibitem{17}  N. Takemoto and A. Becker, J. Chem. Phys. \textbf{134,}
074309 (2011).
\bibitem{18} X. Y. Lai, Q. Y. Cai and M. S. Zhan, New. J. Phys. \textbf{11,} 113035 (2009).
\bibitem{19} A. S. Sanz, B. B. Augstein, J. Wu and C. Figueira de Morisson Faria
(submitted for publication), pre-print arXiv:1205.5298
\bibitem{20} J. Wu, A. S. Sanz, B. B. Augstein and C. Figueira de
Morisson Faria(submitted for publication), pre-print arXiv:
1301.1916
\bibitem{21} Y. Song, F. M. Guo, S. Y. Li, J. G. Chen, G. Chen and Y. J. Yang, Phys. Rev. A \textbf{86,} 033424
(2012).

\bibitem{22} M. Lein, N. Hay, R. Velotta, J. P. Marangos and P. L. Knight, Phys. Rev. A, \textbf{66,} 023805
(2002).

\bibitem{23} G. L. Kamta and A. D. Bandrauk, Phys. Rev. A
\textbf{71,} 053407 (2005).

\bibitem{24} T. Kanai, S. Minemoto, and H. Sakai, Nature
\textbf{435,} 470 (2005).

\bibitem{25} Y. Han and L. B. Madsen, J. Phys. B \textbf{43,}
225601(2010).
\bibitem{26} A. Etches, M. B. Gaarde, and L. B. Madsen, Phys. Rev. A
\textbf{84,} 023418 (2001).

\bibitem{27} Y. Wu, J. T. Zhang, H. L. Ye, and Z. Z. Xu, Phys. Rev.
A \textbf{83,} 023417 (2011).
\bibitem{28} Y. C. Han and L. B. Madsen, Phys. Rev. A \textbf{87,}
043404 (2013).



\bibitem{29} D. Bohm, Phys. Rev \textbf{85,} 166 (1952).
\bibitem{30} D. Bohm, Phys. Rev \textbf{85,} 180 (1952).
\bibitem{31} R. E. Wyatt 2005 Quantum Dynamics with Trajectories (NewYork:
Springer).

\bibitem{32} Y. J. Yang, G. Chen, J. G. Chen and Q. R. Zhu, Chin. Phys. Lett. \textbf{21,} 652 (2004).

\bibitem{33} Y. J. Yang, J. G. Chen, F. P. Chi and Q. R. Zhu, H. X. Zhang and J. Z. Sun, Chin. Phys. Lett. \textbf{24,} 1537 (2007).

\bibitem{34} J. Wang, G. Chen, F. M. Guo, S. Y. Li, J. G. Chen and
Y. J. Yang, Chin. Phys. B \textbf{22,}033203(2013).

\bibitem{35} R. D. Cowan 1981 The Theory of Atomic Structure and Spectra (Univ of California Pr).
\bibitem{36} J. G. Chen, S. L. Zeng and Y. J. Yang, Phys. Rev. A. \textbf{82,} 043401 (2010).
\bibitem{37} J. G. Chen, Y. J. Yang, S. L. Zeng and H. Q. Liang, Phys. Rev. A \textbf{83,} 023401 (2011).





\end{thebibliography}

\begin{figure}[htb]
\includegraphics[width=15cm,height=12cm]{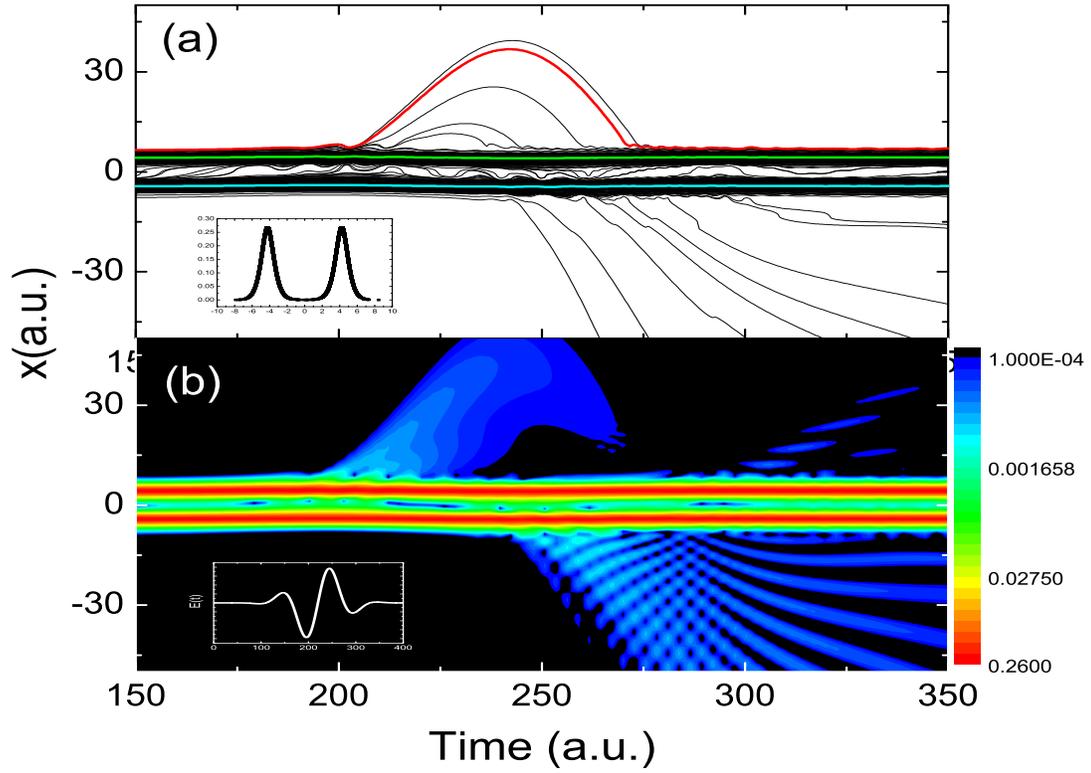}
\caption{\label{fig_1} (Color online) (a) Time-volution of BTs;
(b) time-volution of the electronic probability density and
selected BTs BT(N) and BT(P).}
\end{figure}

\begin{figure}[htb]
\includegraphics[width=15cm,height=10cm]{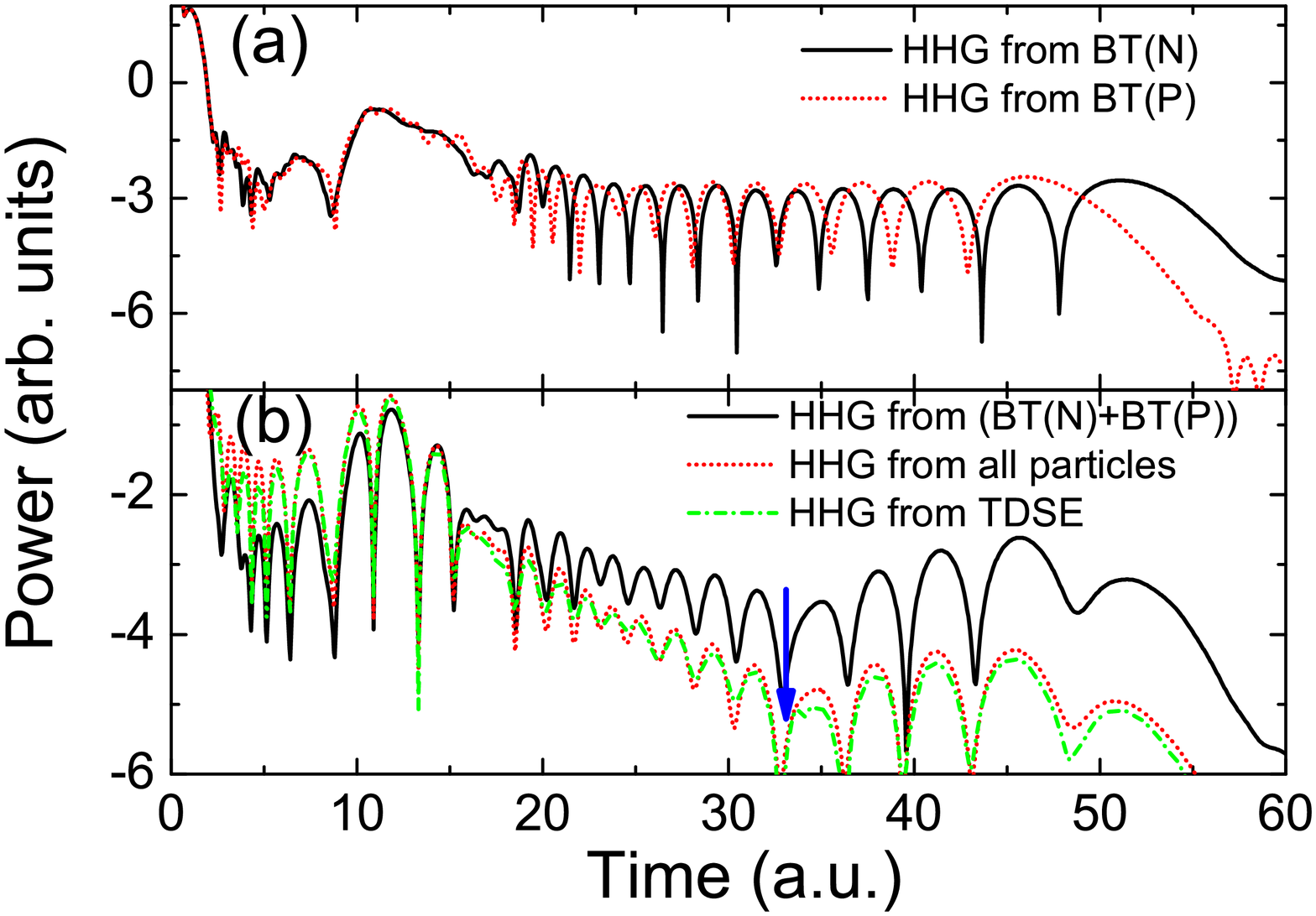}
\caption{\label{fig_2} (Color online) Harmonic spectra from the
diatomic molecule ion irradiated by the driving laser pulse, whose
parameters are $E_0 = 0.1$ and $\omega= 0.057$: (a) calculated
from the BT(N) and BT(P); (b) calculated from the TDSE and
coherent sum of two or all Bohmian particles.}
\end{figure}

\begin{figure}[htb]
\includegraphics[width=15cm,height=15cm]{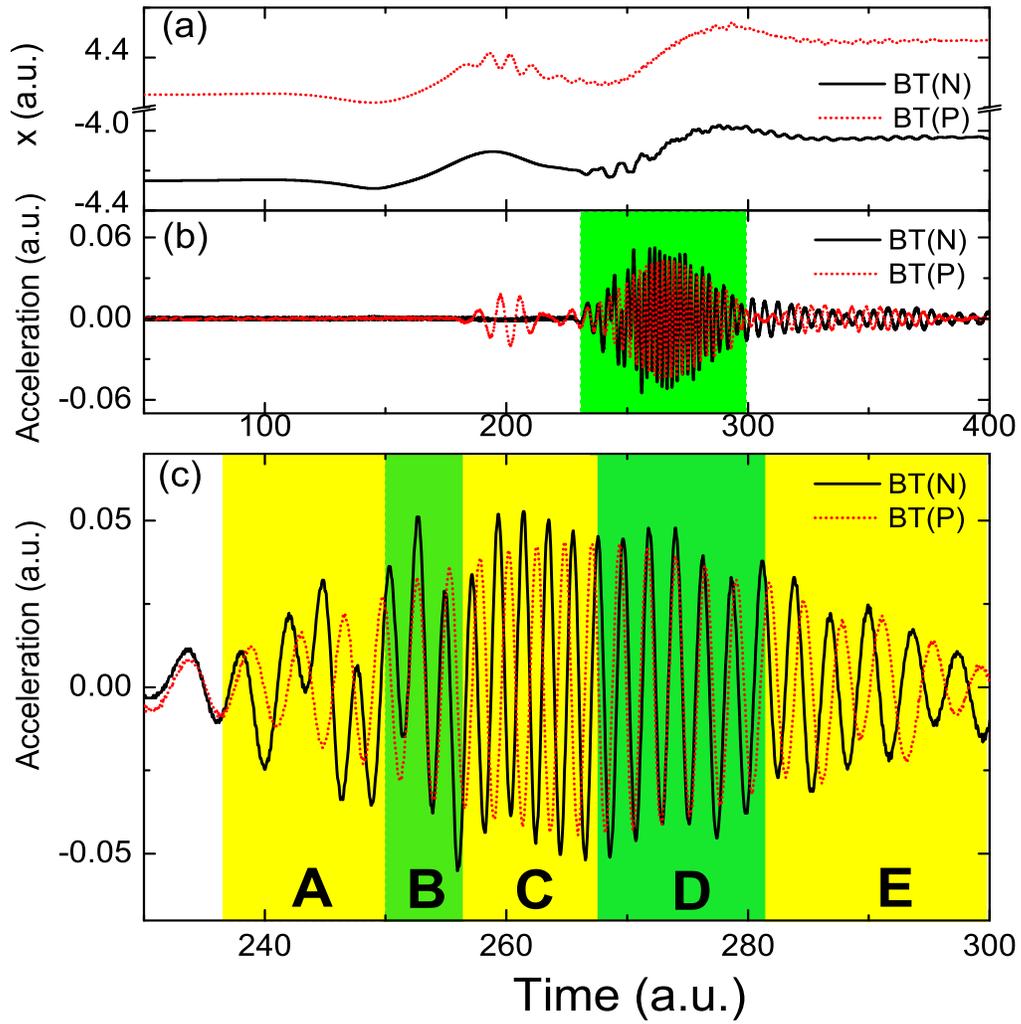}
\caption{\label{fig_3} (Color online) Evolution of BTs BT(N) and
BT(P): (a) time-dependent trajectories; (b) acceleration; (c)
detailed acceleration.}
\end{figure}

\begin{figure}[htb]
\includegraphics[width=15cm,height=15cm]{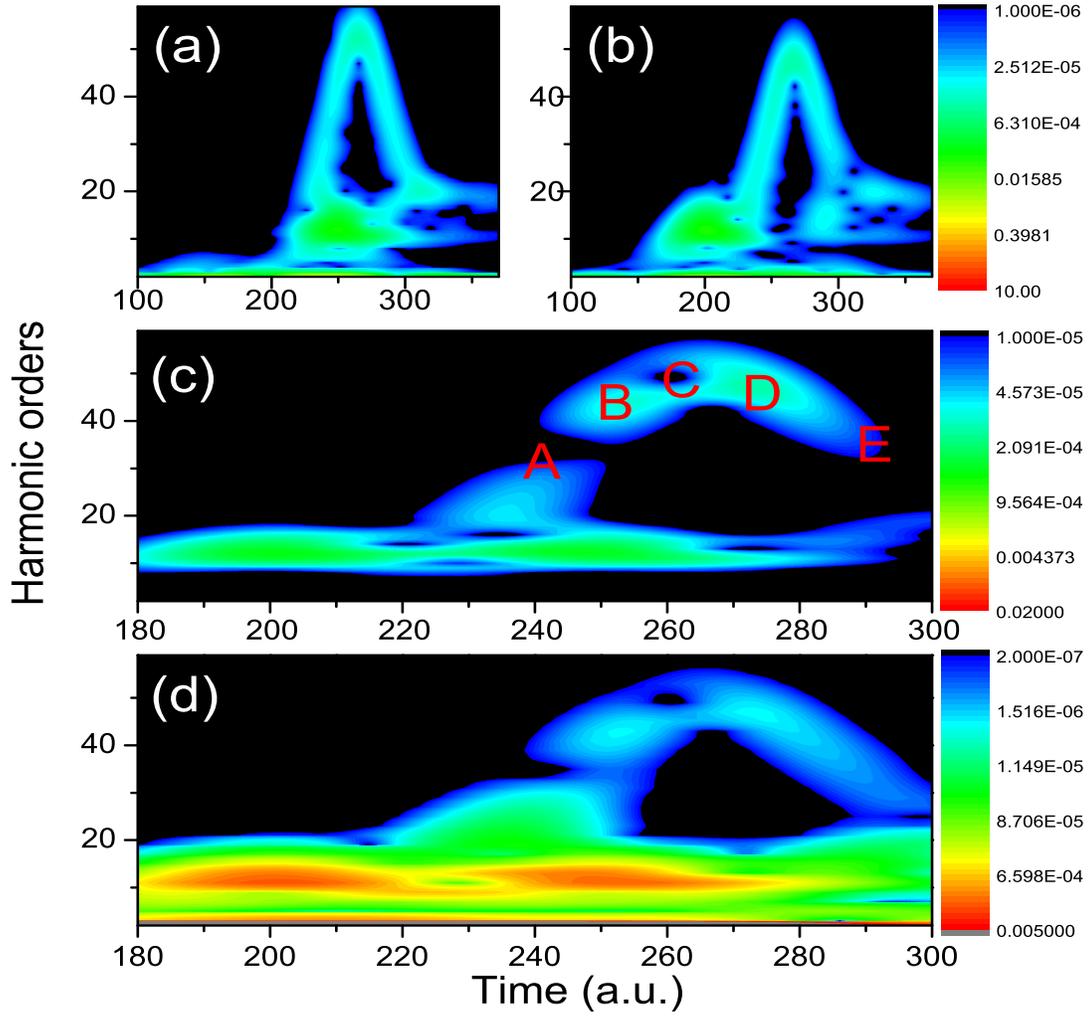}
\caption{\label{fig_4} (Color online) Time-frequency behavior of
harmonic emission calculated from:  (a) BT(N); (b) BT(P); (c)
coherent sum of BT(P) and BT(N); (d) TDSE.}
\end{figure}

\end{document}